\documentclass[10pt]{article}

\usepackage{graphicx,amsmath,amssymb,ragged2e,natbib}

\usepackage[margin=2cm]{geometry}

\title{Spatial dynamics of complex urban systems within an evolutionary theory frame} 

\date{}

\author{
Juste Raimbault\textsuperscript{1,2,3*},
Denise Pumain\textsuperscript{3}\bigskip\\
\textsuperscript{1} Centre for Advanced Spatial Analysis, UCL, London, United Kingdom
\\
\textsuperscript{2} UPS CNRS 3611 ISC-PIF, Paris, France
\\
\textsuperscript{3} UMR CNRS 8504 G{\'e}ographie-cit{\'e}s, Paris, France\bigskip\\
* \texttt{juste.raimbault@polytechnique.edu}
}

\begin{document}

\maketitle

\begin{abstract}
This chapter is about Complexity and Spatial Dynamics in Urban Systems. Strong inequalities in the size of cities and the apparent difficulty of limiting their growth raise practical issues for spatial planning. At a time when new constraints in terms of limited energy and raw material resources or possible catastrophic events such as pandemics are challenging further urban expansion, it is important to consolidate the theories from various scientific disciplines to estimate to what extent the urban dynamics can be modified. While briefly reviewing the contributions to urban theories provided by the new developments in complexity sciences, we first advocate for the soundness of urban theories. Second, we develop our original approach considering spatial interaction and evolutionary path dependence as major features in the general behavior of urban entities. Third, we test these principles grounded in an evolutionary theory of urban systems by experimenting four dynamic models of urban growth calibrated on harmonized empirical data sets with comparisons across the whole world.\medskip

\end{abstract}

\justify

\section{Introduction}

The context of this chapter, namely ``Entropy, Complexity, and Spatial Dynamics: The Rebirth of Theory'' is challenging. The problem is twofold: first, it means demonstrating how the insights provided by concepts and models associated with the notion of complexity have contributed to theories on spatial dynamics; second, it questions whether this is a rebirth of the theoretical approach in this field, which would therefore presume its relative abandon before.

This chapter provides answers to these two questions about a particular dimension of spatial dynamics, that of cities and systems of cities. In response to the second question, and considering the progress of research on urban complexity over the last forty years or so, we identify an enrichment and consolidation of existing theories by the principles of complex systems rather than a true renaissance of urban theory. Indeed, contrary to what some scholars claim, the theories developed around the spatial dynamics of urbanization processes and the evolution of settlement systems are far from being obsolete \citep{brenner2014urban}. Neither are they limited in their fundamentals and application to only one part of the world \citep{robinson2016comparative}. \cite{scott2015nature} and \cite{wu2020emerging} rightly criticized both ``theories''. We may add that many scholars did not abandon the former urban theories but on contrary tested, revised and completed them during the last four decades \citep{pumain1997pour,pumain1998urban,pumain2003approche,pumain2020theories,batty2013new}. The major theoretical challenge of that period was three folds: first, to shift from idiographic postures to nomothetic ones; second, from static views toward dynamic and evolutionary; and third, to really transfer concepts and models from natural sciences to build a socially relevant knowledge. Elements of knowledge from ancient theories can consolidate when situated in their geo-historical contexts, when tests of these theoretical propositions are carried out on subsets of empirical data that are well identified in time and space, and when the means are provided to articulate the observations made at multiple levels of geographic scales.

Starting from such a spiral theoretical conception of the cumulativeness of knowledge \citep{pumain2005cumulativite}, we try here to demonstrate how it is possible to construct an evolutionary theory of cities and systems of cities. This theory is abductive in the sense that it is constructed by frequent to-and-fro movements between empirical observations, logical propositions and mathematical and computer models. This theory is not entirely new as it integrates elements of knowledge already well identified by specialists in urban issues, such as geographers, economists, historians and archaeologists, sociologists, town planners and architects. Its originality stems from progress in international comparisons made through using new harmonized databases, methods for validating the results of computer simulation models, and putting these results into the perspective of complex systems theories.

This chapter brings a step forward in the long process of building and testing an evolutionary theory of systems of cities. This theory is based on empirical observations and tested with dynamic models that are designed for simulating urban development at various spatial and temporal scales. We summarize here the methodology and results already obtained in the GeoDiverCity project \citep{pumain2015multilevel,cura2017old,pumain2017urban}, including USA, Europe and BRICS countries, and the work of \cite{raimbault2020empowering} completing them with new datasets at world scale and other types of models. These models are designed for explaining urban growth and city size distributions with an increasing deepening in the complexity of the implemented processes \cite{cottineau2015growing}. They are all conceived for exploring the correspondences between urban trajectories observed at the meso-level of individual cities and the structuring of systems of cities at macro geographical scale. The validity and representativeness of these four models of complex urban systems are tested and their variations across regions of the world within different spatial frames are studied \citep{raimbault2018calibration,raimbault2020indirect}. This work is made possible by using different data sources, namely the GeoDivercity database and a new data source that for the first time compares the evolution of cities globally over a period between 1970 and 2015, the GHSL database. Eric \cite{denis2020more} already explored the source statistically for comparing urban sprawl trends in the countries of the world. \cite{raimbault2020empowering} calibrate four different models of urban growth on both sets of data, across different regions of the world, and analyze their results and assess for each their possible contribution to the theory of urban complex systems. We recall here the structure of these models, in relation to the question of how to build integrated urban theories.

The rest of this chapter is organized as follows: we first develop the importance of constructing urban theories and how they relate to models; we then develop the details of urban dynamics models introduced in the context of the evolutionary urban theory; we finally discuss important open problems crucial in the development of such theories.

\section{For a theoretical rearming in urban science}

To clarify our position vis-à-vis urban theories, we propose a first discussion of the diversity of possible meanings attributed to this word among the disciplines that have been interested in explaining the evolution of cities. This discussion is based on a few recent publications, without claiming to be exhaustive within the limits of this chapter. The main question we address is to understand how the disciplines most formerly interested in the scientific object ``city'' have been able to revise or complete their theoretical statements using the contributions proposed by the natural sciences, which have been advancing interesting concepts and models for thinking about and studying complex systems for some forty years. To begin with, we can start with the definition of complexity as proposed by Aura \cite{reggiani2009complexity} for whom ``the term "complexity" embeds both the assemblage of different units in a system and their intertwined dynamics. In other words, the term "complexity" is strictly related to the concept of networks''. In this definition we would emphasize the physical and societal interactions that are creating and using these networks with a consequence of making the evolution of ``different units'' strongly interdependent on each other. That leads to a specific deepening in the concept of co-evolution for cities belonging to systems of cities \citep{paulus2004coevolution,raimbault2020unveiling}.

Two repeated observations prompted the development of a theory of the evolution of systems of cities. The first, empirical, is the long persistence of urban networks in large integrated territories, which retain, over time, the same spatial configurations of the relative size of cities. This is visible for example on maps representing the population of cities in proportional circles more than a century apart, in Europe, in India, or even in the United States since 1950. The second results of applying to the evolution of these urban population distribution a simple stochastic statistical model formalized by the French statistician Gibrat since 1931. This model states that the growth of urban populations is proportional to the initial size of cities during each short time interval, with fluctuations such as growth rates (amount reported to the population) are statistically independent from the size of the cities and from one time interval to the next. The interest of this model of spatially distributed urban growth in integrated systems of cities is to predict that the result of such a growth process is always a lognormal distribution. Thus, the model provides a first statistical explanation to the ``mystery'' of the Zipf's law still recently mentioned by the economist Paul \cite{krugman1996confronting}. A huge literature is dedicated to Zipf’s law and Pareto distributions that are observed in many natural and social systems, to the extent that these ``universal'' laws, together with fractal spatial patterns, have been sometimes considered as the ``signature'' of complexity in different systems. The underlying regularities in the urban spatial dynamics that these models capture have attracted many scientists from various disciplines to propose different theories of urban complexity.

\subsection{Which acceptation of ``theory'' for complex urban systems?}

A preprint posted on the Internet in January 2020 \citep{lobo2020urban} deserves particular attention at first sight for three reasons. First, it announces an ``integrated theory''. Second, it covers the full historical period ``from the first cities to sustainable metropolises''. Third, 35 names of prominent scholars signed it, which is rather rare among papers dealing with objects of social sciences.

On which propositions do they come all in agreement? A first one is about the necessary multidisciplinary character of ``urban science'', which we fully agree with as well. A second proposal is about acknowledging the fact that cities are the product of a long but relatively recent and socially driven historical evolution, which should lead to the conclusion that urban theories are part of social sciences dealing with historical objects. However, the paper acknowledges, ``a basic tension in current urban science between approaches from the social sciences and those from the natural sciences'' (p.3). The authors conclude nevertheless: ``the time is ripe for a targeted integration of the social and physical sciences of cities and urbanism'' (p.4).

The author’s argumentation about this point becomes ambiguous and even contradictory. When they enunciate what the urban theory should be, they suggest Darwinian biology as a model of science to imitate in focusing on processes rather than forms, which we can agree. But they add: ``A theoretically grounded science of cities should capture fundamental processes that lie at the core of all human spatial agglomerations, whether these are past or present, agrarian or industrial, in developed economies or developing countries'' (p. 11). They clearly tend to think of universal ``fundamental principles'' that would appear as the core of urban science, while the categories that were elaborated and coined by history or other disciplines of social sciences would be considered as part of a “context”. Should we not require that a “historically grounded” urban science would have processes from societal theories as ``fundamental principles''? In other words, the processes leading to the elaboration of concepts such as ``agrarian'' or ``industrial'' society or ``developing country'' may receive at least the same level of attention that the one of ``agglomeration'' if we want to understand urban evolution and properly integrate the disciplinary points of view.

That is why we are not sure that the ``scaling analytic framework'' is indeed the first thing to mention of what we have learned from ``urban science injected into the study of cities and urbanization'' – even if that may appear paradoxical speaking from a geographer’s point of view, always caring about scale! We can nevertheless subscribe to most of the authors’s 13 propositions enabling them to converge toward ``unified perspectives'' –although a few would deserve more discussion. As the authors, we are convinced that there is a need for urban theory and that is it necessarily grounded in history, properly tested with abundant and diverse data compared in space and time. Of course, we agree to consider cities as places of high density, frequent social interaction and mixed activities, knowing that their size is both a cause and consequence of their creativity, acknowledging the universal character of urban hierarchies and the specific role of innovations in their development. However, we could question the list of nine propositions that the authors consider as new insights provided by the new ``urban science''. The list of 23 questions to solve in future research is not so new either. We acknowledge that the sciences of complex systems that developed during the second half of 20th century can be granted for having challenged social sciences toward more interest in mathematical and computational formalisms and having provided a series of formalized tools and models and a common vocabulary that established a bridge between researchers in mathematics and physics and social sciences. Nevertheless, the urban theory cannot be reduced to its quantifiable dimensions only.

We may wonder if any ``integrated theory'' is yet imaginable because at a given moment in science, and in social sciences especially, there are a plurality of theories. Social sciences may deserve a specific definition of complexity that would depend on the number of different disciplines that are required for providing a satisfying explanation or interpretation of a particular object, by borrowing concepts and models from these disciplines, each having investigated a category of complex processes, to make the object intelligible at a particular granularity level of description. Thus it is likely that while we pretend currently to build an evolutionary theory of urban systems, through integrating as much as possible from accumulated empirical knowledge in geographical, archeological and historical tradition, using statistical and simulation models from the new ``urban science'', other fields are not yet satisfied with a theory of urban hierarchy, such as \cite{krugman1996confronting} for whom it remains a ``mystery'' that cannot be properly derived from the principles of economic theory only.

\subsection{Minimum requirement for a theory of urban complex systems}

When adopting a nomothetic attitude in urban research, we adopt a series of epistemological concepts and practices that are common with other sciences. The starting point is to work with \textit{empirical data that are properly defined before being properly measured}. Taking care for meaningful definitions of cities is not a trivial exercise, because it supposes to enter and to understand the criteria that societies chose for establishing a distinction between rural and urban settlements \citep{rozenblat2020extending}. It is obvious for any social scientist that such a distinction has a political origin, even if many state governments in the last two centuries decided to objectivize as much as possible their definition through elaborating statistics. For centuries being ``urban'' was part of a social status, an attribute conveyed to people rather than to places. Most social scientists are uneasy with attempts at qualifying the urban areas delineated from satellite images of built-up areas or street networks as ``natural cities'' \citep{jiang2015zipf}. Of course, cities belong to ``nature'' in using material resources and energy and hosting humans. They may even become greener if they succeed in managing the next ecological transition... However, social organizations are driving cities' generative processes, rather than natural features. There are traces of the history of cities as political constructions in the social representations opposing peasants and urban citizens, with sometimes practical and economic consequences on their way of life, if you think for instance of the former Chinese hukou system \citep{wu2020emerging}. There are traces as well in the decisions of naming ``urban'' different places such as India distinguishing ``Statutory towns'' and ``Census towns'' \citep{swerts2018diffuse}. Some countries periodically delineate the perimeters of urban areas to follow the spatial expansion of their cities, such as Germany or Russia, whereas others do not.

This did not prevent too many experimented and incidentally prominent scholars to capture “urban data” without any care about their quality, selection, claiming and striking results after crunching them with sophisticated statistical or mathematical models. Such an attitude is detrimental, for instance regarding the endless controversies about the models of city sizes distribution, for building a sound science that accumulates knowledge with reproducible means \citep{pumain2012theorie}. The resulting cacophony is sometimes disentangled in courageous articles, such as for instance \citep{cottineau2017metazipf,cottineau2020metametazipf} in a meta-review of hundreds of references, but certainly represents a waste of time and efforts that could be avoided. A recent paper seemed promising in announcing the observation of worldwide trends in urbanization from 1950 to 2030 using some 1857 cities larger than 300 000 inhabitants from 155 countries \citep{egidi2020long}. The paper offers surprising results such as a systematic ``inverse relationship'' between urban growth and city size in all regions ``up to the late 1990s''. At first sight, this could result from mixing countries with opposite urban growth trends, but the authors observe it as well in homogenous sub regions. However, after questioning the authors who kindly accepted to precise the origin of their data, it seems that the selected cities were delineated over time in a way not measuring the spatial expansion of the urban area. This could explain why the results seem in contradiction with those obtained from large parts of the world during the same period with harmonized urban databases \citep{cura2017old}. It also reminds about the necessity of starting with a sound conceptual definition of cities for measuring their attributes and controlling the possible effects of the data selection on the results. The urban definition may vary but should be in accordance with concepts and theory, i.e. with the research question and the selection of attributes that are measured for answering them \citep{rozenblat2020extending}.

To specify the fundamentals of our theory we retain a conceptual framework that relies on the long history of social sciences investigating cities and interpreting them \citep{pumain1996theoriser}. Thus even if we can integrate insights from complex systems science such as fractals and scaling laws, we do not agree that all fundamentals should stem from ``physics for society'' \citep{caldarelli2018physics}. While maintaining a strong attention to the spatial dimension, our concern with space is neither static nor geometric. It is geographical, and geo-historical. Space is crucial in the definition of cities and systems of cities, it is a social space whose properties are revised and updated with the changing technologies, and it is as well made of transformed natural space in the process of using earth resources and developing political territories. The dimensionality of cities is constrained by the accessible space for face-to-face daily activities, whereas transportation networks enabling less frequent encounters may generate and sustain interdependencies at much longer distances leading to identify systems of cities. At each observation level, those of individuals (persons or institutions), cities and systems of cities, new attributes and new concepts do correspond to emergent phenomena in complex systems that are produced from multiple interactions.

Although the recent development of networks and the accelerating circulation of information seem to blur the three-level conceptual definition of these geographical objects, we believe that this representation remains useful and relevant for guiding urban research and applications. An interesting contribution by C. \cite{roth2006reconstruction} discusses the emergentist view compared to the reductionist and suggests that different modes of access and not only bottom- up interactions could be part of the generative processes of the features constituting one level – which is the view point from statistical mechanics in envisaging self-organizing systems. This rejoins our theory for which urban interactions shaping cities and systems of cities are mostly but not exclusively conceived as emanating from one low level toward the higher level, in societal processes that take a long time, often longer than a human life. Over decades and sometimes centuries, most frequent interactions among urban stakeholders at micro level shape the emerging properties of a city at meso-geographical level, and inter-urban interactions generate the hierarchical organization and the functional differentiation of cities within the system of cities at macro-level. However, for instance, a location decision taken by an individual firm (at micro-level) can operate a definite change in the relative specialization of a particular city within the system of cities (at macro level), as well as many policy regulations decided at macro-level may directly change the situation of an individual firm or citizen at micro-level inside a city.

Sociologists such as Michel \cite{grossetti2020matiere} discuss the ontology of entities that scholars identify in social sciences. A science of cities is possible because cities are remarkably persistent entities, among all societal institutions entities of the meso level, in an intermediary position between individual agents and the territories that embed them. Through their interactions however, they build at a much higher level, comparable to the one of civilizations, spatial organizations that we name systems of cities and maintain their ``emerging'' properties over much longer durations. The rhythm of relative societal change within cities is much slower than the changes in human life and change in systems of cities is much slower than change in individual cities. Our theory of cities as complex social systems embedded within systems of cities and territories is \emph{evolutionary}. However, the historical evolution is different from biological, of course because its processes are social, meaning they are guided with human intentions (even if that causality is not always directly effective) and they proceed much more rapidly than changes in biology.

While trying to identify major processes guiding the evolution of cities within system of cities, we stressed the role of information and innovation \citep{lane2009complexity}. Innovation when defined as a socially accepted invention may represent a form of social organization or a belief or a practice as well as a technological device or a new service. We shall develop further in the models presented below how innovation and its more or less always hierarchical diffusion among cities within systems of cities appears as a major dynamic process for explaining the expansion of urbanization and the persistency of cities’ relative situation in urban systems.

\subsection{Defining integrated theories}

An issue worth mentioning before developing further the relation between theories and models is the definition of a theory itself, and furthermore of an ``integrated theory''. \citep{lobo2020urban} use this term several times without detailing its exact meaning - to what extent ambiguity can be harmful in this aspect? The nature, structure and function of scientific theories has always been one core subject in philosophy of science. According to \cite{winther2016structure}, three main views are currently accepted for the structure of scientific theories: (i) the syntactic view which sees theories as mathematical logical constructions; (ii) the semantic view, which considers a set-theoretic approach and understands theories as set of models, the theory being itself a model; and (iii) the pragmatic view, which includes more complex ways of knowledge production and components, including informal or implicit knowledge such as analogies, but also taking into account how knowledge is socially produced and organized. Following \cite{suppe2000understanding}, the semantic view was proposed to overcome limitations of the syntactic view (such as the fact that theories are not axiomatic systems). It was suggested by \citep{halvorson2012scientific} that the semantic view is also limited to be able to grasp the complexity of scientific knowledge production. One stream of ideas within the pragmatic view is the pragmatic view of models, focusing on the diversity and functions of models. In that context, the epistemological positioning of scientific perspectivism \citep{giere2010scientific} advocates for a plurality of scientific perspectives, of which models are media. According to \citep{callebaut2012scientific}, this approach is highly relevant to deal with complex and multi-scalar problems. In the context of urban theories, \citep{pumain2020conclusion} showed a plurality of theories and models in practice. We therefore suggest that a pragmatic view of theory is the most relevant to build integrated theories, which will have to couple multiple heterogeneous components. This aspect also comes into play when considering tools or methods used to support the theory. Although it is not clearly stated, it seemed implied by \citep{lobo2020urban} that an integrated theory will have to be ``rigorous'' and formulated mainly with mathematics. While theories in physics almost necessarily include mathematical formalism and derivations, this is not the case in social science, and knowledge can be communicated and formalised through other ways, as recalled by \cite{banos2013pour} as one of the crucial principles for modeling and simulation in social science.

Finally, the meaning of ``integrated'' is a rather open question. There are many terms to distinguish between different degrees of interaction between disciplines, from multidisciplinarity (combining approaches from different disciplines) to actual interdisciplinarity which implies that new knowledge is created beyond the disciplines \citep{huutoniemi2010analyzing}. There is also no clear definition of model coupling in the literature, nor of theory coupling. Model coupling can range from a ``loose coupling'' where outputs of one model are used as inputs of another \citep{clarke1998loose}, to stronger couplings where both model dynamics influence each other in time or even where additional coupling mechanisms are constructed and implemented, leading to new hybrid models \citep{mustafa2017coupling}. Furthermore, there may exist different types of integration: in the roadmap for the study of complex systems, \citep{chavalarias2009french} propose a two-dimensional reading of integration, through the construction of integrated disciplines spanning across one field at multiple scales (vertical integration), and the investigation of transversal fundamental questions that are pervasive to different types of complex systems (horizontal integration).

We postulate here that ``integration'' refers to such strength of model and theory coupling, but the question of its proper definition and quantification remains open. We will in the following build on possible interactions between models and theories, leading to the position that coupling models is a possible way to build integrated theories.

\subsection{How to make models and theory dialogue?}

If the theory is constructed from empirical observations, prioritized and simplified into stylized facts, the design of the model should enable to reproduce these stylized facts. Dynamic models start from an initial observed situation that is described in a simplified way by state variables (such as the population of cities, their income level or their employment structure). These models reconstruct the evolution of these quantities on the basis of mechanisms (rules and parameters), which represent the interactions (i.e. social processes) that are supposed to explain the transformations of the state variables over time \citep{pumain2013theoretical}.

For a model to be an interesting tool for testing a theory, a correspondence must be established between the way it is implemented in the model and the expected result of the simulation. Selecting the stylized facts to be reconstructed and the granularity of their description is a crucial choice for the success of the exercise. For example, it is rather pointless to give the model the sole objective of correctly simulating a distribution of city sizes, in the form of a lognormal or Pareto law. This form of distribution (described as an 'over-identified model' by B. \cite{robson1973urban}) can indeed emerge from a whole series of mathematical or computer mechanisms, some of which have little to do with the functioning of a system of cities. Without aiming for absolute realism but in order to get closer to it and make better use of modelling, it is necessary, for example, to go one-step further and ensure that the growth process produced by the model corresponds in detail to what is summarized in \cite{gibrat1931inegalites}'s statistical model \citep{modica2017methodological}.

All models described below include the well-established fact that the urban hierarchy receives a relevant statistical explanation from Gibrat's model of urban growth. However, this random mathematical model has at its core a hypothesis that makes it little credible to found a theory of cities, since it assumes that all cities (elements of the statistical distribution) are independent of each other, as well as the random events that lead to their growth. Such a hypothesis enters in conflict with the definition, the functioning of cities and the processes of their emergence and their evolution, which organizes them into systems of co-evolution, counteracting these simplifying assumptions a priori. In fact, the slight deviations noted in relation to the hypotheses of the model have led to the completion of other hypotheses that adapt this simple model, powerful but too general, to different historical and geographical contexts.

In the Simpop family of models, the reconstruction of cities' trajectories is based on their interactions using the multi-agents systems as computing technique for simulation. This technique was chosen because, when compared to mathematical systems of differential equations, they enable more flexible representation of the variety of spatial interactions that characterize inter-urban trade according to the functional specialization of cities. In brief, the so-called ``central functions'' providing services to the population interact with their market according to gravitational principles, while manufacturing and touristic cities select sometimes distant but specific places to interact with, and administrative functions deal in a systematic way with all other cities in their circumscription. More detailed precisions can be injected in the simulation, as for instance the political selection of some urban places for developing industries of national interest during the socialist regime in the former Soviet Union \citep{cottineau2014evolution}.

So, logically the theory is as much as possible embedded in the model, but how can the model contribute to the building of the theory? Of course, the usual recursive processes during model building may ensure that the major theoretical principles are correctly framed in the implemented model. Nevertheless, until recently there was too little confidence in the results obtained from simulation models when using multi-agents modelling techniques \citep{rey2015plateforme}. The validation procedures were considered as not sufficiently reliable for guaranteeing that the estimated parameters were the only values capable of generating the required features of the system’s evolution and that all the included rules were necessary to produce the simulated result. A real scientific breakthrough for SSH occurred when the combined use of genetic algorithms and distributed computing enabled to shift from the usual few hundred simulations of the same model toward several hundred millions \citep{schmitt2015half}. This experiment not only provides an almost full exploration of the parameter space but also ensures that all parameters leading to the expected result are both sufficient and necessary \citep{reuillon2015new,raimbault2019methods}. Although large-scale computing has always been anchored within quantitative geography practices \citep{rey2019calcul}, these results were particularly novel in the way models, methods and theory were intertwined.

This process of an iterative construction of theories and models corresponds to the latest developments of an ``abductive'' epistemology, which succeeded historically to deductive approaches in classical geography and purely inductive approaches \citep{varenne2018theories}. In that context, the relation between theories and models can be understood through the lens of knowledge domains \citep{raimbault2017applied}: scientific knowledge production is achieved through perspectives which rely fundamentally on the model as their main medium, but are also composed of other types of knowledge: theoretical (defined in this case in the pragmatic way), empirical (facts derived from data), data itself, methods, and tools implementing methods. This typology may not be exhausted but already captures the strong interaction between different components of knowledge, and how the progressive iteration between their contents, including models and theories, is central. We see here that a third dimension of integration could be proposed, as the integration between knowledge domains, accounting for how strong are the dependencies between models and theories for example. \citep{raimbault2020relating} has proposed a similar view as the core of an ``applied perspectivism'', an epistemological stance aimed at expliciting disciplinary positions through reflexivity and enhancing interdisciplinarity. In that context, a way forward to couple theories and build integrated theories would be through the coupling of models themselves. In that context, model validation methods are key, and these can be concretely diffused into diverse disciplines with interdisciplinary and reflexive knowledge dissemination practices \citep{leclaire2019retour}. The approach of applied perspectivism is based on a strong interaction between models and theories, and as we put it above, on the use of new model exploration and validation methods.

\section{Simulation models for systems of cities}

Following what we just developed, simulation models and their systematic exploration seem to be a powerful way to (re-)build urban theories. We propose now to detail the structure of a few simulation models for systems of cities, which were all introduced within the frame of the evolutionary urban theory. We describe in particular the processes taken into account by each model, and what the corresponding exploration results may imply for the theory.

The different models we describe below were benchmarked and applied to several systems of cities worldwide by \citep{raimbault2020empowering}. They rely on the same fundamental assumptions and share a common basic structure and formulation: (i) agents are cities, characterized a main state variable which is population; (ii) building on the Gibrat model, these add to endogenous growth additional processes which account for space and spatial interactions; (iii) they simulate not stochastic distributions of city sizes, but their average in time, and are thus specified on these averages only. The interaction between cities, which could be included in the covariance structure for a fully stochastic model, is in that case captured by the spatial interaction terms.

\subsection{A model developing urban hierarchy with transportation network}

One dimension of urban systems which is expected to significantly influence system trajectories is physical infrastructure, and more particularly transportation infrastructure \citep{dupuy1987vers}. This aspect is included in a model of spatial interaction taking into account physical transportation networks, studied by \citep{raimbault2020indirect}. It was extended into a co- evolution model between cities and abstract transportation networks by \citep{raimbault2020modeling}, and refined into a co-evolution model taking into account an explicit spatial structure of the network by \citep{raimbault2020hierarchy}. The population of cities $P_i$ are updated at each time $t$ following

\begin{equation}
	P_i (t + \Delta t) = P_i(t) \cdot \left[ 1 + \Delta t \cdot \left( r_0 + \hat{w_G} \cdot \sum_{i \neq j} \left( \frac{P_i(t)P_j(t)}{P(t)^2}\right)^{\gamma_G}  \cdot \exp \left( - \frac{d_{ij}}{d_G} \right)\right) \right]	
\end{equation}

with $\Delta t$ time step, $r_0$ fixed endogenous growth rate, $\hat{w_G}$ relative share of spatial interactions (the parameter is in practice renormalized by the average of interaction potentials), $P(t)$ total population, $\gamma_G$ hierarchy of spatial interactions, $d_G$ range of spatial interactions, and $d_{ij}$ geographical or network distance (yielding two different versions of the model). The initial model of (Raimbault, 2018b) includes an additional term accounting for centrality effects in the network, with feedback processes of flows going through a city on its population growth rate. This model is completed by a network evolution module to give the co-evolution model: network link effective speeds are updated at each time by

\begin{equation}
	d_l(t+ \Delta t) = d_l(t) \cdot \left[ 1 + \Delta t \cdot g_M \left( \frac{1 - \left(\frac{\varphi_l}{\varphi_0}\right)^{\gamma_N}}{1 + \left(\frac{\varphi_l}{\varphi_0}\right)^{\gamma_N}} \right) \right]	
\end{equation}

where $d_l(t)$ is the length of link $l$, $g_M$ a maximal speed increase parameter, $\varphi_l$ the flow within the link, $\varphi_0$ the threshold parameter (above which link speed increase) determined in practice through a quantile parameter $\varphi_0^{(q)}$, and $\gamma_N$ the hierarchy of the process. Flows within a physical network can be determined by any network assignment algorithm, given the spatial interaction flows obtained at the previous step.

\begin{figure}
	\includegraphics[width=\linewidth]{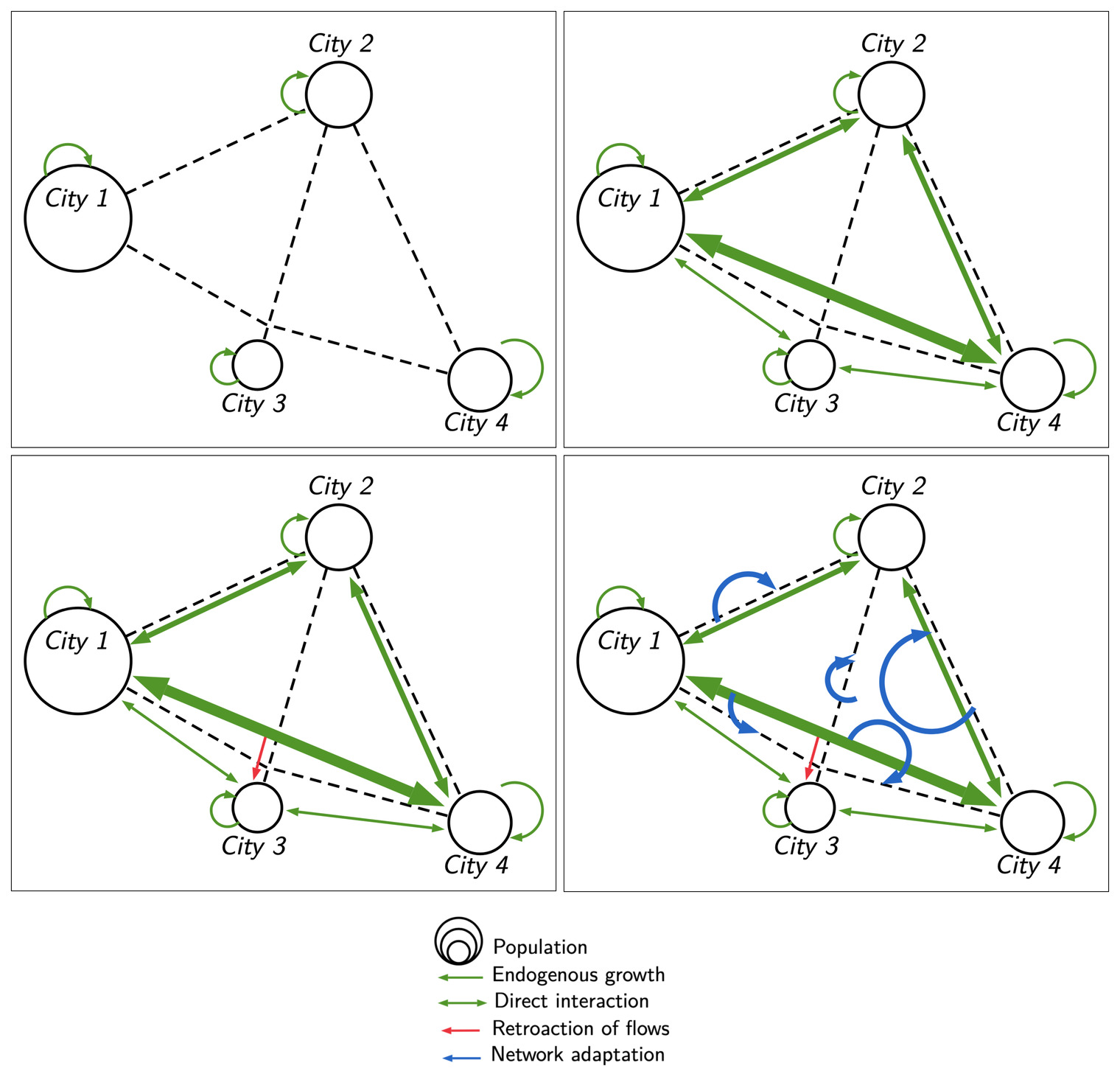}
	\caption{Schematic representation of the different processes included in the macroscopic co- evolution model for cities and transportation networks. Cities are characterized by their population (circle area) and are linked by a transportation network (dashed lines). City population grows endogenously (Gibrat model, first panel), through spatial interactions (second panel) and feedback of flows (third panel). Network links evolve according to the flow they carry (fourth panel).\label{fig:fig1}}	
\end{figure}

The complete model workflow is synthesized in Fig.~\ref{fig:fig1}. This summarizes the superposition of the different processes linked to cities and the transportation network. The exploration and application of this model can bring new insights for theory building, including for example: (i) do network-mediated spatial interactions explain well urban trajectories? - it was shown by \citep{raimbault2020indirect} that in terms of fitting error for the French system of cities, the feedback of network flows actually improved model performance, implying that these network processes are good candidates as explanatory variables for urban dynamics; (ii) can co- evolution in the statistical and population sense be effectively reproduced between cities and transportation infrastructures? - \citep{raimbault2020modeling} exhibits a large variety of co-evolution regimes produced by the model; (iii) what are the common dynamics of urban and network hierarchies? - \citep{raimbault2020hierarchy} investigated this question and showed much richer dynamics than a simple self-reinforcement. These results suggest that next theoretical developments should not neglect infrastructure networks, the role of co-evolution, and to rethink concepts of hierarchy in systems of cities.

\subsection{A model explaining urban growth with economic and environmental processes}

Economic processes are the subject of large parts of the literature studying urban dynamics, for example in urban economics, economic geography, or regional science, but not only. The MARIUS model family developed by \citep{cottineau2014evolution} is based on economic exchanges between cities, within the frame of the evolutionary urban theory. Following \citep{cottineau2015modular}, the model family is composed of a baseline model capturing spatial interactions (the Gibrat mechanism is not included in this model which can however been seen as having the same basis as other models presented here, where the endogenous growth mechanism has been deactivated), and incremental mechanisms which are a growth bonus for cities with more interactions, a fixed cost of transactions, geographical effects such as the local presence of resources and policies with a territorial impact, and the possibility to modulate urbanisation speed with external temporal constraints. We describe here only the baseline model, reformulated by replacing the power spatial interaction function with an exponential, for the sake of simplicity. Given population of cities at the initial step $P_i(0)$, an abstract wealth state variables is constructing following a scaling law by taking $W_i(0)=P_i^{\alpha_W}$, where $\alpha_W$ is a model parameter. At each time step, supply and demand functions are computed as scaling laws of the current population for each city, potential flows are computed with spatial interactions, balance of exchanges and wealths are updated accordingly. This gives the following non-linear spatial interaction model to update city wealth:

\begin{multline}
	W_i (t + \Delta t) = W_i (t) + w_0 \cdot \sum_j \exp \left( - \frac{d_{ij}}{d_G} \right)	\cdot \left( \min \left[ \frac{P_i^{\alpha_S} P_j^{\alpha_D}}{\sum_k P_k^{\alpha_S} \exp(-d_{ij}/d_G)} ; \frac{P_i^{\alpha_D} P_j^{\alpha_S}}{\sum_k P_k^{\alpha_D} \exp(-d_{ij}/d_G)} \right]\right. \\ \left. - \min \left[ \frac{P_i^{\alpha_D} P_j^{\alpha_S}}{\sum_k P_k^{\alpha_S} \exp(-d_{ij}/d_G)} ; \frac{P_i^{\alpha_S} P_j^{\alpha_D}}{\sum_k P_k^{\alpha_D} \exp(-d_{ij}/d_G)} \right] \right)
\end{multline}

where $w_0$ is a normalization parameter, $d_G$ the spatial interaction range, $\alpha_S$ the supply scaling parameter, and $\alpha_D$ the demand supply scaling parameter. The population increment is then computed as a function of the scaled wealth increment, as

\begin{equation}
	P_i (t+\Delta t) = P_i (t) + \frac{1}{w_0} \left( W_i (t+\Delta t)^{\alpha_P} - W_i (t)^{\alpha_P} \right)	
\end{equation}

where $\alpha_P$ is the population scaling parameter. More detailed model explanation, exploration, and application to the Former Soviet Union can be found in \citep{cottineau2014evolution,cottineau2015growing,cottineau2015modular}.

Some insights for theory from this model family are for example detailed by \citep{cottineau2015modular}: (i) which mechanisms among the different economic processes included are better at explaining urban trajectories, when taking into account parsimony, i.e. considering only models with the baseline and one other mechanism? - different optimal model structure were found depending on the calibration structure, suggesting a strong break in stationarity due to the fall the Soviet Union; applying a similar approach to other systems of cities remains difficult as the model was tailored to that case, but would bring significant insight into the question of non-stationarity of urban trajectories; (ii) when considering all processes, which are the most important to explain city trajectories? - a surrogate statistical model applied on values of the calibration error as a function of parameters, for an optimal model population composed of a fixed number of individual for each model instance, gives the relative contribution of each mechanism to the fit improvement; (iii) what are the properties of cities on which the model does not perform well? - this provides feedback on the specific theoretical assumptions and mechanism survey done during the construction of the model family for this particular case study.

\subsection{A model linking urban growth with innovation as spatial diffusion process}

\subsubsection{The Favaro-Pumain model}

The spatial diffusion of innovations has been coined as a fundamental process driving urban dynamics \citep{hagerstrand1968innovation}. In economics, many approaches consider innovation as crucial, such as in endogenous growth theory \citep{aghion1998endogenous}. The theoretical and empirical study of innovation clusters and knowledge spillovers is also a considerable stream in economic literature \citep{audretsch2005does}. In the context of the evolutionary urban theory, the SimpopLocal model \citep{schmitt2014modelisation} proposes to link the diffusion of innovation with the emergence of cities in an agent-based approach. Based on similar ideas, but within a discrete dynamical system formulation which inspired the formulation of other models we gave here,  \cite{favaro2011gibrat} introduced an extension of the Gibrat model considering spatial interactions for the dynamics of populations and for the diffusion of innovations. We describe here the model as specified by \citep{raimbault2020model}. In this model, cities are characterized by their population $P_i(t)$ and by a level of adoption of a set of innovations $\delta_{c,i,t}$ corresponding to proportions such that $\sum_c \delta_{c,i,t} = 1$. Populations are updated following a spatial interaction model, where attractivity of cities is modulated by the local level of adoption relative to the global level, by

\begin{equation}
	P_i (t+\Delta t) = P_i(t) \cdot \left[1 + \Delta t \cdot \left(r_0 + \hat{w_I} \cdot \sum_{i\neq j} \frac{P_i(t) P_j(t)}{P(t)^2} \cdot \exp \left(- \frac{d_{ij}}{d_G} \right) \cdot \prod_c \delta_{c,i,t}^{\varphi_{c,t}} \right) \right]
\end{equation}

The parameter $\hat{w_I}$ is in practice normalized by the average of potentials. The global adoption level is defined by $\varphi_{c,t} = \sum_i \delta_{i,c,t} P_i / \sum_{i,c} \delta$. Parameters $r_0$ and $d_G$ are as previously the endogenous growth rate and spatial interaction range. Innovations are diffused between cities also with a spatial interaction model, according to

\begin{equation}
	\delta_{c,i,t+\Delta t} = \frac{\sum_j p_{c,j,t}^{1/u_c}\cdot \exp\left(-d_{ij}/d_I\right)}{\sum_c\sum_j p_{c,j,t}^{1/u_c}\cdot \exp\left(-d_{ij}/d_I\right)}	
\end{equation}

where $p_{c,i,t} = \delta_{c,i,t}\cdot \frac{P_i}{P}$ is the population share in city $i$ adopting innovation $c$, and $u_c$ is the innovation utility (innovations with a higher utility will diffuse faster). The last mechanism in a model step corresponds to the introduction of new innovations - we describe it in the next subsection which corresponds to a modified version of the model tailored to capture urban evolution.

\subsubsection{An explicit model of urban evolution}

\begin{figure}
	\includegraphics[width=\linewidth]{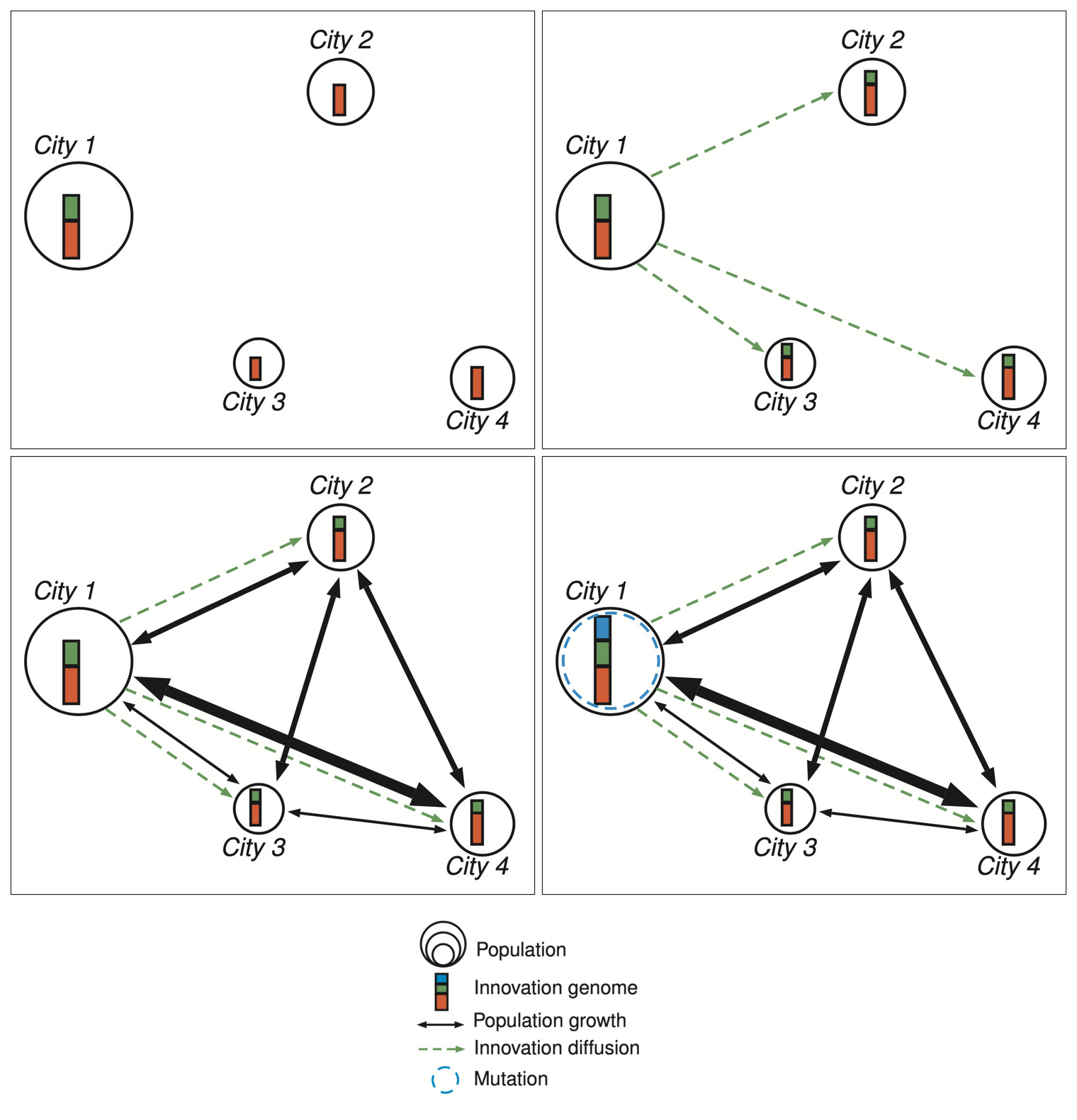}	
	\caption{Schematic representation of the different processes included in the innovation diffusion urban evolution model. Cities are characterized by their population (circle area) and their innovation shares or genome (color boxes). Innovations are diffused between cities with a spatial interaction model (second panel). City sizes are updated according to spatial interactions and attractivity linked to the share of the dominant innovation (third panel). Genomes are mutated with possibly the introduction of new innovations (fourth panel).\label{fig:fig2}}
\end{figure}

The initial Favaro-Pumain model described by \citep{favaro2011gibrat} is an urban dynamics model, but not an urban evolution model in a sense that would extend biological and cultural evolution \citep{mesoudi2017pursuing}. To have evolution in a population, one must have fundamental processes of transmission and transformation, leading to differentiating sub- populations. Elementary units of transmission are genes in biology, while the concept of meme was introduced in cultural evolution. In what would consist elementary units and processes of an urban evolution in that sense remains an open question, but \citep{raimbault2020model} proposed a possible instantiation, extending the Favaro-Pumain model. Elementary units are innovation, and the urban genome corresponds to the innovation proportions $\delta_{c,i,t}$. Transmission processes correspond to the spatial diffusion of innovations and the evolution of city sizes due to spatial interactions, as a kind of spatial crossover. Transformation processes are achieved through local mutations of the genome, corresponding to random discoveries of new innovations. In the initial model, the stylized fact of creation-destruction was implemented by deterministically inserting a new innovation when a certain adoption threshold was attained, and innovation utility was multiplied by a fixed factor. In the urban evolution model, at each time step each city has a probability to innovate given by

\begin{equation}
	p = \beta \cdot \left( \frac{P_i(t)}{\max_k P_k(t)} \right)^{\alpha_I}	
\end{equation}

where $\beta$ is an intrinsic innovativity parameter, and the probability to innovate is proportional to a scaling function of size with exponent $\alpha_I$. The utility of a new innovation is drawn stochastically from a distribution with average the current empirical mean of existing innovation and with standard deviation a fixed parameter $\sigma_U$. The new innovation has a penetration share of $\delta_0$ in the new city, which genome is modified by rescaling the share of other innovations. A schematic summary of the innovation diffusion urban evolution model is shown in Fig.~\ref{fig:fig2}.

Insights for the construction of theory that can be drawn from this model are diverse. First, with the Favaro-Pumain model only, one can investigate to what extent hierarchical diffusion of innovation occurs and successive waves influence urban growth. When considering the urban evolution model, the possibility for the model to lead to the emergence of effective co- evolution niches (i.e. regions of space diverging significantly in terms of genome composition in time) is an important feature for the role of co-evolution in the urban theory. Results of the exploration by \citep{raimbault2020model} were preliminary and did not investigate specific indicators quantifying co-evolution, but found intermediate spatial interaction ranges leading to structures similar to innovation clusters. Finally, perspectives to parametrize this model with empirical innovation data (using patents as proxy for example) can be linked to feedback on the effective role of innovation diffusion in urban dynamics.

\section{Discussion}

We have previously discussed the importance of theory (whatever its meaning) in urban studies, and its relation to modeling. In particular, we recalled how models and theories can dialogue to strengthen conclusions on both sides, and advance towards the construction of integrated theories. We finally illustrated with concrete urban simulation models diverse ways theory can be instantiated into simulation models and how the exploration, validation and application of these models can inform back the theory. Many obstacles still remain in that context, and we discuss now two particular issues that in our opinion will require a particular attention to reach effectively integrated theories.

\subsection{Multi-modeling and concurrent hypotheses}

One significant difficulty, particularly linked to the integration of complementary/concurrent hypotheses or theories, is how to properly compare them in terms of relevance to the integrated knowledge built. To tackle it, we discuss above how \cite{cottineau2015modular} followed the procedure of multi-modeling, in the sense of a full model in which incremental mechanisms can be included or not. Combining specific technical tools such as automatic code generation and methodological tools with a dedicated niched genetic algorithm to calibrate a large set of these models, they were able to produce a population of optimal solution and evaluate statistically the contribution of diverse mechanisms to model fit, yielding therein a benchmark of mechanisms in terms of explanatory power. This approach is however possible only when all models can be integrated into a single framework and directly compared. In the case of \citep{raimbault2020empowering}, which benchmarked simple versions of the models we described above, such automatic generation and niched calibration could not be achieved, and models were independently calibrated on two objectives and the final Pareto fronts compared. This was still possible as the models have the same input and output data and a similar structure. How to couple and integrate models when they are highly heterogeneous, at different scales and on totally different objects? How to build multi-models of heterogeneous models, ideally automatically towards the idea of ``model crushing'' coined by \citep{openshaw1993modelling}? These questions remain rather open, although initiatives such as the model exploration platform OpenMOLE are built towards the facilitation of tackling such issues \citep{raimbault2019methods}. Another methodological difficulty in model comparison that should not be neglected is the fair comparison of models, i.e. taking into account overfitting and keeping parsimonious models. Generic information criteria for simulation models that would render a fair comparison possible as Akaike Information Criterion does for statistical models, also remains an open problem \citep{raimbault2020indirect}. These questions directly relate to the nature of integration, of model coupling, and the construction of multi-scale models.

\subsection{Between synthetic and real systems of cities}

One classical mantra in French geography is the tension between ``the general and the particular'' \citep{durand1991particulier}. This opposition can directly be found at the core of theories for systems of cities: what are the general and robust stylized facts that can be attributed to systems of cities in general, and what are the peculiar geo-historical conditions that lead to a particular trajectory. Beyond the myth of ``universality'' advocated by physicists which would imply universal laws derived from the same microscopic processes and which would assume some kind of ergodicity in the system \citep{pumain2012urban}, robust stylized facts such as Zipf and Gibrat laws are well identified in a first approximation. In that context and within the interaction between models and theories, being able to distinguish between intrinsic model dynamics, robust effects due to space that one observe on similar geographical structure, and effects due to the peculiar initial conditions and geo-historical conditions, is a difficult problem. \citep{raimbault2019space} introduced a novel methodological pipeline for sensitivity analysis of spatial models, which enables to test the sensitivity of model outcomes to spatial initial conditions. This requires generating spatial synthetic data that resembles real configurations, in this case population grids at the scale of urban areas. \citep{raimbault2020scala} develop generators at other scales (microscopic building configurations, macroscopic systems of cities, transportation networks). Such work is the basis towards a better understanding on how urban dynamics models behave on synthetic systems of cities compared to real case studies. This is essential for the construction of integrated theories, to understand their level of generality, and how they relate to specific geographical situations.

\section{Conclusion}

We have recalled here the meaning and importance of theories to study complex urban systems, and how a diversity of theories is needed to grasp the multi-dimensional nature of such systems. This diversity can be the basis of coupling between theories, and the construction of integrated theories. What such integration consists of and how to reach it remains in our opinion an open question, which is nevertheless partly answered by the use of simulation models, their coupling and a dynamical dialogue between models and theories. We illustrated these concepts with three urban dynamics models developed within the frame of an evolutionary theory of cities. Many obstacles remain on the path towards an evidence-based integrated knowledge of complex urban systems, but we believe elementary bricks already exist and methodological ways forward are progressively developed.

\end{document}